\documentclass[epj]{svjour}

\usepackage{graphics}
\usepackage{epsfig}
\usepackage{dcolumn}  
\usepackage{bm}  
\usepackage{epsfig}

\usepackage{amsmath}
\usepackage{amssymb}

\begin{document}

\title{Sample-to-sample torque fluctuations in a system of coaxial randomly charged surfaces}

\author{Ali Naji\inst{1,2}\thanks{\email{a.naji@damtp.cam.ac.uk} (corresponding author)} \and Jalal Sarabadani\inst{3} \and David S. Dean \inst{4,5} \and Rudolf Podgornik\inst{6,7,8}}

\institute{                    
School of Physics, Institute for Research in
Fundamental Sciences (IPM), Tehran 19395-5531, Iran \and
Department of Applied Mathematics and Theoretical
Physics, Centre for Mathematical Sciences, University of
Cambridge, Cambridge CB3 0WA, United Kingdom \and
Department of Physics, University of Isfahan, Isfahan 81746, Iran \and
Laboratoire de Physique Th\'eorique (IRSAMC),Universit\'e de Toulouse, UPS and CNRS,  F-31062 Toulouse, France \and
Universit\'e de  Bordeaux and CNRS, Laboratoire Ondes et Mati\`ere d'Aquitaine (LOMA), UMR 5798, F-33400 Talence, France \and
 Department of Theoretical Physics, J. Stefan Institute, SI-1000 Ljubljana, Slovenia \and
 Department of Physics, Faculty of Mathematics and Physics, University of Ljubljana, SI-1000 Ljubljana, Slovenia \and
 Laboratoire de Physique Th\'eorique (IRSAMC),Universit\'e de Toulouse, UPS and CNRS,  F-31062 Toulouse, France
}

\date{Received: date / Revised version: date}

\abstract{
Polarizable randomly charged dielectric objects have been recently shown to exhibit long-range lateral and normal 
interaction forces even when they are effectively net neutral. These forces stem from an interplay between the quenched statistics of random
charges and the induced dielectric image charges. This type of interaction has recently been evoked to interpret  measurements of Casimir
forces {\em in vacuo}, where a precise analysis of such disorder-induced effects appears to be necessary.  Here we consider the
torque acting on a randomly charged dielectric surface (or a sphere) mounted on a central axle next to another randomly charged surface
and show that although the resultant mean torque is zero, its sample-to-sample fluctuation exhibits a long-range behavior
with the separation distance between the juxtaposed surfaces and that, in particular, its root-mean-square value scales with 
the total area of the surfaces. Therefore, the 
disorder-induced torque between two randomly charged surfaces is expected to be much more pronounced than the disorder-induced lateral force 
and may provide an  effective  way to determine possible disorder effects in experiments,  in a manner that is  independent of the usual normal force measurement. 
}

\maketitle

\section{Introduction}

The surfaces of dielectrics, crystalline solids and metals often exhibit random monopolar charge distributions \cite{science11,barrett,kim}. Random charges may 
result from adsorption of contaminants and/or the presence of impurities that can generate surface charges which depend strongly  on the method of preparation 
of the samples. Surface charge disorder may also originate from the variation of the local crystallographic axes of the exposed surface of a clean 
polycrystalline sample which induces  a variation of the local surface potential \cite{barrett,speake,kim}. The heterogeneous structure of the charge disorder 
on dielectric surfaces and thus its statistical properties can be determined directly from Kelvin force microscopy measurements \cite{science11}. Randomly
charged surfaces are equally abundant in colloidal and soft matter systems   \cite{Rudi-Ali1,Rudi-Ali2,Rudi-Ali3}, examples arise in surfactant coated surfaces \cite{surf}, unstructured proteins
and random polyelectrolytes and polyampholytes \cite{ranpol}. 

A number of authors \cite{kim} have pointed out that disorder effects may significantly influence the measurement of the Casimir-van der Waals (vdW) forces between solid surfaces in vacuum. These forces act between all objects and are relatively short-ranged in nature. Recent ultrahigh sensitivity experiments of the Casimir force have however revealed a 
residual long-range interaction force which dominates at sufficiently large separations,
and it has been suggested that it is due to surface disorder effects \cite{kim}.

Recently it has been proposed that quenched random charge disorder on surfaces as well as in the bulk of dielectric slabs can lead to long-range  interactions even when the surfaces are 
net-neutral \cite{cd1,cd2}. These long-range interactions stem from a subtle interplay
between the quenched statistics of surface or bulk charges and the image charge effects 
generated by the dielectric discontinuities present at the bounding surfaces. 

It was  subsequently demonstrated \cite{dean2011} that two randomly charged surfaces can interact with both random {\em normal} forces, whose mean value turns out to be {\em finite} and long-ranged as noted above, 
and  random {\em lateral} forces, which--for two juxtaposed planar surfaces carrying statistically homogeneous random charges--turn out to be {\em zero} on the average. Both 
quantities however show sample-to-sample fluctuations whose root-mean-square value
also exhibits a long-range behavior with the separation distance between the surfaces. 

In this study, 
we  pursue our analysis of disorder effects on long-range interactions  and investigate the
torque acting on a randomly charged planar or (spherical) dielectric object mounted on a central axle next to a 
randomly charged dielectric substrate. We show that although the resultant mean torque in this system is zero, its sample-to-sample fluctuation 
exhibits a long-range behavior with the separation distance. Even more importantly,  the 
root-mean-square value of the torque fluctuations scales with 
the total area of the surfaces and thus represents an extensive quantity.  Therefore, the 
disorder-induced torque between two randomly charged surfaces is expected to be much more pronounced than 
the disorder-induced lateral force 
and may present a more effective method to quantify charge disorder effects in experiments. 
Our results on disorder induced torque may also be related to the so-called lock and key phenomena, which underpin highly specific interactions between complex biological molecules such as proteins, where long-range electrostatic interactions can induce pre-alignment which enables complex molecules to interact in a biologically useful manner \cite{molec}.

\section{Two plane-parallel dielectric slabs}

\begin{figure}[t]
\begin{center}
\vspace{-5cm}
\centerline{\includegraphics[angle=0,width=8.5cm]{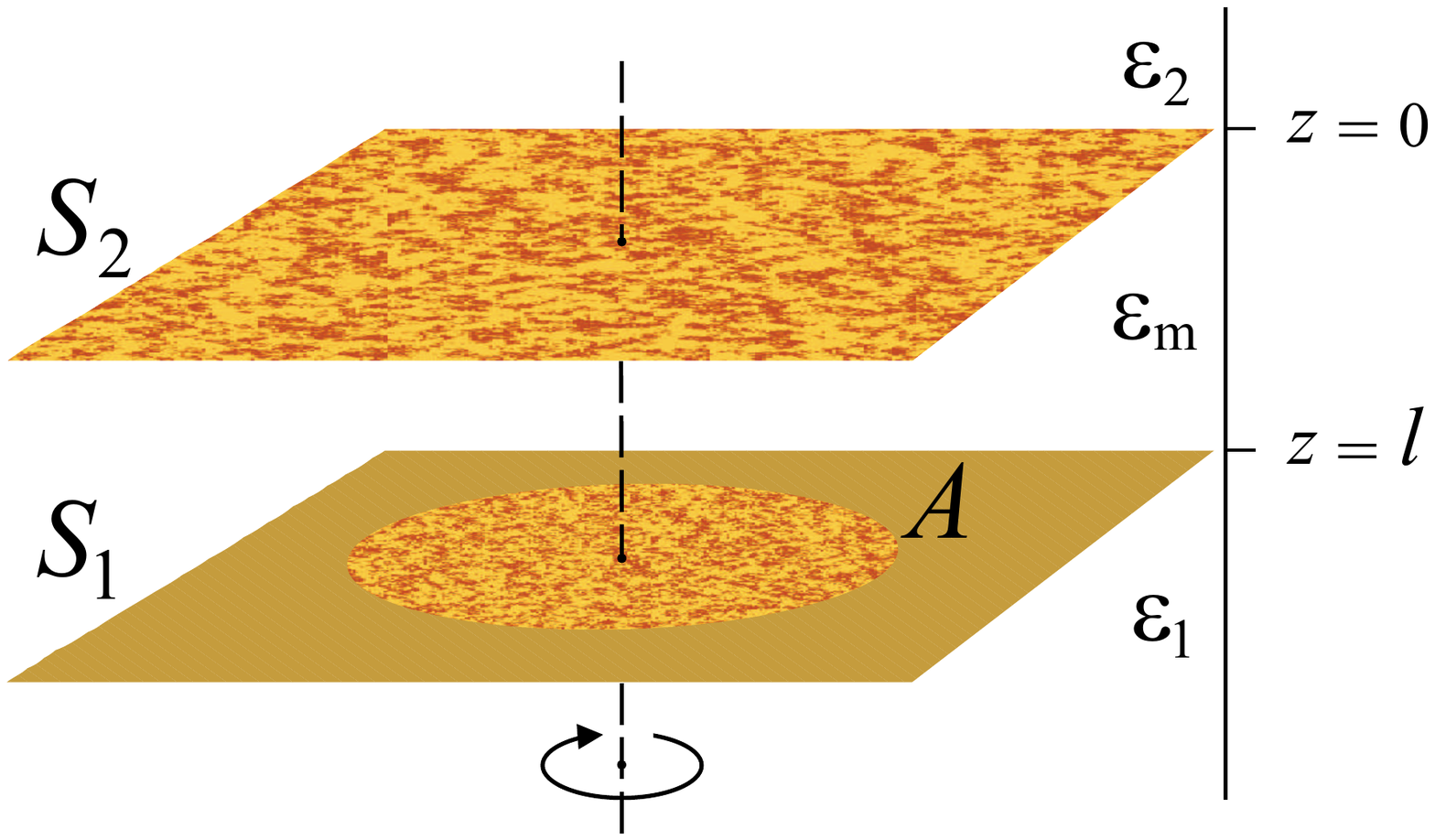}}
\vspace{-2cm}
\caption{(Color online) Schematic representation of two dielectric half-spaces 
at a separation distance of $l$
carrying quenched random charges  (shown by small dark and light brown patches) on their bounding surfaces. The charges on the lower surface are assumed to be distributed in a circular region of area $A$, which is then taken to infinity. The lower surface 
is allowed to rotate around the central $z$ axis as shown in the figure.}
\label{fig:fig_pp_schematic}
\vspace{-.8cm}
\end{center}
\end{figure}

Consider two plane-parallel dielectric half-spaces 
with the bounding 
surfaces separated 
in the $z$
direction by a distance $l$.  The 
surface 
at $z=0$
belongs to a 
dielectric half-space with the dielectric constant $\varepsilon_2$ and the 
surface 
at $z=l$ belongs to a dielectric half-space with the
dielectric constant $\varepsilon_1$ (see Fig. \ref{fig:fig_pp_schematic}). We call these 
surfaces $S_2$ and $S_1$, respectively. We denote by $\varepsilon_m$ the dielectric constant of the intervening material. Let each 
surface (labelled by $\alpha, \beta=1, 2$)
have a random surface charge density $\rho_\alpha({\bf x})= \rho_ \alpha({\bf r}, z) $  with zero mean (i.e., the surfaces are {\em net-neutral}) 
and the correlation  function in the  plane of the slabs  (${\bf r}, {\bf r}'\in S_1, S_2$)
\begin{equation}
\langle \rho_ \alpha({\bf r}, z) \rho_\beta({\bf r}', z')\rangle=\delta_{\alpha\beta}\,  g_{\alpha s}\,\delta(z-l_\alpha)
\delta(z'-l_\beta)\,C_\alpha({\bf r}-{\bf r}'),
\end{equation}
where we define $l_2=0$ and $l_1=l$. In addition, we  assume that the charge distribution on 
surface $S_1$ is restricted to a finite area $A$. The 
surface $S_1$ is assumed to be mounted on an axle that allows for rotation
around its central symmetry axis. In the case where 
the random charge is made up of point charges of signs $\pm e$ of surface density 
$n_{\alpha s}$, we may write the variance of the  charge disorder as $ g_{\alpha s} = e^2  n_{\alpha s}$. The correlation function $C({\bf r}-{\bf r'})$ has dimensions of inverse length squared, meaning that its two dimensional Fourier transform is dimensionless. Typically the  values of $n_s$ for reasonably pure samples are smaller than the bulk disorder variance which is usually in the  range of between $10^{-11}$ to $10^{-6}~ {\rm nm}^{3}$ (corresponding to impurity charge densities of $10^{10}$ to $10^{15}$ $e/\rm{cm}^3$ \cite{kao,pitaevskii,cd1,cd2}).

The electrostatic energy of the system is given by
\begin{equation}
E = \frac{1}{2}\int d{\bf x} \,\phi({\bf x})\rho({\bf x})
\end{equation}
where $\rho({\bf x})$ is the total charge density and 
\begin{equation}
\phi({\bf x})= \int d{\bf y} \,G({\bf x},{\bf y})\rho({\bf y})
\end{equation}
 is the electrostatic potential, 
while $G({\bf x},{\bf y})$ is the Green's function obeying
\begin{equation}
\varepsilon_0\nabla\cdot[\varepsilon({\bf x}) \nabla G({\bf x},{\bf y})] =-\delta({\bf x}-{\bf y})
\end{equation}
with $\varepsilon({\bf x})$ being the local dielectric function. Upon changing the charge distribution the 
corresponding change in the energy of the system is thus given by
\begin{equation}
\delta E = \int d{\bf x} \,d{\bf y}\,\delta\rho({\bf x}) G({\bf x},{\bf y})\rho({\bf y}).
\end{equation}
If the charge distribution on the surface $S_1$, $\rho_1({\bf x})$, is made up of point charges,  we have 
\begin{equation}
\rho_1({\bf x}) = \sum_{n\in S_1}q_n \delta({\bf x}-{\bf x}_n),
\end{equation}
where $q_n$ is the charge at the site ${\bf x}_n$.
Now on rotating the 
surface $S_1$ by an angle $\theta$ around its symmetry axis, that is to say in the direction perpendicular to  the normal between the bounding surfaces of the two dielectric media, we find that the new charge distribution is given by
\begin{equation}
\rho'_1({\bf x}) = \sum_{n\in S_1}q_n \delta({\bf r}-\hat R_\theta\, {\bf r}_n)\delta(z-z_n), 
\end{equation}
where $\hat R_\theta$ is the two-dimensional rotation matrix. For an infinitesimal rotation angle $\delta\theta$, one has 
$\hat R_{\delta\theta} = 1- \imath (\delta\theta)\,  \hat \sigma_2$, 
where $ \hat \sigma_2=\left( {\begin{array}{cc} 0 & -\imath  \\ \imath & 0  \end{array} } \right)$ is the Pauli matrix. 
This means that we can write (assuming the summation over the in-plane Cartesian components $i, j = 1, 2$ and using the fact that the diagonal elements of $\hat \sigma_2$ are zero)
\begin{eqnarray}
&&\delta\rho({\bf x}) = \delta\rho_1({\bf x}) = \imath(\delta\theta)\,  (\hat \sigma_2)_{ij} \sum_{n\in S_1}q_n ({\bf r}_n)_j\qquad \qquad\qquad\nonumber\\
   &&\quad\times \frac{\partial}{\partial r_i}\delta({\bf r}-{\bf r}_n)\delta(z-z_n)
= \imath(\delta\theta)\,  (\hat \sigma_2)_{ij} r_j \frac{\partial}{\partial r_i}  \rho_1({\bf x}).  
\end{eqnarray}

As the surface $S_1$ is rotated the self-interaction between the charges on each surface is
unchanged, thus the energy change is only given by the interaction of the charges and image
charges in $S_1$ with those in $S_2$. We may thus write
\begin{eqnarray}
\delta E &= &  \imath(\delta\theta)\,  (\hat \sigma_2)_{ij}\int 
d{\bf r'}d{\bf r}\,dz\,dz'\,\big[ r'_j \frac{\partial}{\partial r'_i}  \rho_1({\bf r}',z')\big] \quad \qquad \nonumber\\
  &&\qquad \qquad \qquad \qquad\quad \times G({\bf r}-{\bf r}';z,z')\rho_2({\bf r},z),
\end{eqnarray}
where ${\bf r}'$ and ${\bf r}$ are again the two-dimensional coordinates in the planes of $S_1$ and $S_2$ respectively and $z$ and $z'$ 
are the respective coordinates normal to the planes. We thus note
that the integration over the coordinate ${\bf r}'$ is over a finite area $A$, while that over ${\bf r}$ is
unrestricted. The torque $\tau$ acting on the surface $S_1$ is thus given by $\delta E = -(\delta \theta)\, \tau$.
As the charge distribution on the surfaces $S_1$ and $S_2$ are uncorrelated we find that 
$\langle \delta E\rangle = -(\delta \theta)\langle\tau\rangle = 0$,
where $\langle\cdots\rangle$ denotes the ensemble average over the random charge distributions. Thus the mean torque is zero.  The variance of the torque is obtained using 
$\big\langle (\delta E)^2\big\rangle = (\delta \theta)^2\langle\tau^2\rangle$ as
\begin{figure*}[t!]
\begin{eqnarray}
\langle \tau^2\rangle 
= &&  - (\hat \sigma_2)_{ij} (\hat \sigma_2)_{mp} \bigg\langle 
\int d{\bf r'}\,d{\bf r}\,dz\,dz'\,d{\bf s'}\,d{\bf s}\,d\zeta\,d\zeta'  \, \big[ r'_j\frac{\partial}{\partial r'_i}\rho_1({\bf r}',z')\big]G({\bf r}-{\bf r}';z,z')\rho_2({\bf r},z) \nonumber\\
   && \qquad\qquad\qquad \qquad\qquad\qquad  \qquad\qquad\qquad  \qquad\qquad\qquad  \times
        \big[ s'_p\frac{\partial}{\partial s'_m} \rho_1({\bf s}',\zeta')\big]G({\bf s}-{\bf s}';\zeta,\zeta')\rho_2({\bf s},\zeta)\bigg\rangle, 
   \label{eq:tau_long}
\end{eqnarray}
\hrule
\end{figure*}
\begin{center}
{\em see equation~(\ref{eq:tau_long})}  
\end{center}
where as before ${\bf r}', {\bf s}'\in S_1$, ${\bf r}, {\bf s}\in S_2$ and the summation is again over the in-plane Cartesian components $i, j, m, p = 1, 2$. Again as the charge distributions on the two surfaces are assumed to be uncorrelated, the  only nonzero correlations in the above are given by
\begin{eqnarray}
\langle \rho_2({\bf r},z)\rho_2({\bf s},\zeta)\rangle &=& g_{2s} \delta(z)\delta(\zeta)C_2({\bf r}-{\bf s}), \\
\langle  \rho_1({\bf r}',z') \rho_1({\bf s}',\zeta')\rangle 
&=& g_{1s}\delta(z'-l)\delta(\zeta'-l)C_1({\bf r}'-{\bf s}').\qquad
\end{eqnarray}
This then yields 
\begin{figure*}[t!]
\begin{equation}
\langle \tau^2\rangle = - (\hat \sigma_2)_{ij} (\hat \sigma_2)_{mp} \, g_{1s}g_{2s}
	\int d{\bf r'}\,d{\bf r}\,d{\bf s'}\,d{\bf s}\, 
		G({\bf r}-{\bf r}';0,l)G({\bf s}-{\bf s}';0,l)C_2({\bf r}-{\bf s})\big[r'_j s'_p \frac{\partial}{\partial r'_i}\frac{\partial}{\partial s'_m} C_1({\bf r}'-{\bf s}')\big].
	   \label{eq:tau_long2}	
\end{equation}
\hrule
\end{figure*}
\begin{center}
{\em see equation~(\ref{eq:tau_long2})}
\end{center}
We now write the above result in terms of the two-dimensional in-plane Fourier transforms  $\tilde G$ and $\tilde C_2$ of the functions $G$ and $C_2$ and carry out the integrations over  the coordinates ${\bf r} $ and ${\bf s}$ of the infinite
plane $S_2$ to obtain 
\begin{eqnarray}
	\langle \tau^2\rangle &=& - (\hat \sigma_2)_{ij} (\hat \sigma_2)_{mp} \, g_{1s}g_{2s}\int \frac{d{\bf k}}{(2\pi)^2}\, \tilde G({\bf k};0,l) \qquad\qquad \nonumber\\
	&&  \qquad\qquad\quad\quad\times \tilde G(-{\bf k};0,l) \tilde C_2(-{\bf k})  ~J_{jpim}({\bf k}).
\end{eqnarray}
Now we have to evaluate the integral over coordinates of $S_1$ which has the form
\begin{equation}
  J_{jpim}({\bf k})  \!= \!\int_{S_1\times S_1} \!\!\!\!\!\!\!\!\!d{\bf r}'d{\bf s}'\big[r'_j s'_p \frac{\partial}{\partial r'_i}\frac{\partial}{\partial s'_m} C_1({\bf r}'-{\bf s}')\big] e^{-\imath{\bf k}\cdot
	({\bf r}'-{\bf s}')}.
\end{equation}
In order to proceed we must assume that the correlation length of the charge disorder is much smaller
than the linear dimensions of the area $A$ on $S_1$ which is covered by random charges.  We introduce the relative coordinate ${\bf u}' = {\bf r}'-{\bf s}'$ 
to obtain
\begin{equation}
  J_{jpim}({\bf k})  \!=\! -\!\!\int_{S_1\times S'_1}  \!\!\!\!\!\!\!\!\! d{\bf r}'d{\bf u}'\,\big[r'_j (r'_p-u'_p) \frac{\partial}{\partial u'_i}\frac{\partial}{\partial u'_m} C_1({\bf u}')\big] e^{-\imath{\bf k}\cdot
	{\bf u}'},
\end{equation}
where $S_1'$ is the shifted integration region over ${\bf u}'$. Now with the assumption that the correlation length of the charge disorder is much smaller than the linear dimensions of the
area $A$ on $S_1$ we
see that the first term above dominates for a large system. Furthermore, we can take the integration over 
${\bf u}'$ to be over ${\mathbb R}^2$ to obtain 
\begin{equation}
J_{jpim} = k_i k_m {\tilde C}_1({\bf k}) K_{jp} 
\end{equation}
where 
\begin{equation}
K_{jp} = \int_{S_1} d{\bf r}\  r_j r_p.
\end{equation}
This then gives the general result
\begin{eqnarray}
	\langle \tau^2\rangle &=& -  \, g_{1s}g_{2s}\int \frac{d{\bf k}}{(2\pi)^2}\, \tilde G({\bf k};0,l)   \tilde G(-{\bf k};0,l)\qquad\qquad \nonumber\\
	  &&\quad\quad \times   \tilde C_2({\bf k})    \tilde C_1({\bf k}) \big[ (\hat \sigma_2)_{ij} (\hat \sigma_2)_{mp} k_i k_m  K_{jp}\big]. 
\end{eqnarray}
This formula can then be applied to a general 
finite two-dimensional charged area $A$ of arbitrary shape assuming that it is sufficiently large. 
An alternative, more geometric form of this result is
\begin{eqnarray}
\langle \tau^2\rangle & =&  g_{1s}g_{2s}\int \frac{d{\bf k}}{(2\pi)^2}\, \tilde G({\bf k};0,l)   \tilde G(-{\bf k};0,l)\qquad\qquad\nonumber\\
  &&\qquad\qquad\quad\quad \times \tilde C_2({\bf k})    \tilde C_1({\bf k}) \int_{S_1} d{\bf r} \,|{\bf k}\times {\bf r}|^2 ,
\end{eqnarray}
where the integration over $S_1$ is about the axis of rotation.
Specializing to the case where the area $A$ on the surface $S_1$ is a disc and the axis of rotation is at its center, we find that
\begin{equation}
K_{jp} = \delta_{jp} \frac{\pi R^4}{4} = \delta_{jp} \frac{A^2}{4\pi},
\end{equation}
where $R$ is the disc radius and $A=\pi R^2$ is its area. 
Now using the fact that $(\hat \sigma_2)_{ij}(\hat \sigma_2)_{mj} = - \delta_{im}$ and that $\tilde G({\bf k})$ and $\tilde C_\alpha({\bf k})$ are functions of $|{\bf k}|=k$ only, 
we may write
\begin{equation}
	\langle \tau^2\rangle = \frac{A^2 g_{1s}g_{2s}}{8\pi^2}\int d k\,  k^3
	     [\tilde G(k;0,l)]^2 \tilde C_2( k)  \tilde C_1( k),
	     \label{eq:pp_tau2}
\end{equation}
which is the final expression of the torque fluctuations in the plane-parallel geometry. 
This result may then be compared with the lateral force fluctuations which were derived in the form \cite{dean2011} 
 \begin{equation}
\langle F_i^{(L)}F_j^{(L)}\rangle\!=\!
 \frac{A\delta_{ij}g_{1s}g_{2s}}{2\pi }\!\!\int \! kdk \ k_i k_j [\tilde G(k;0,l)]^2\tilde C_1(k)\tilde C_2(k), 
 \label{ff1}
 \end{equation}
leading to an intuitively clear physical  relation between the lateral force fluctuations and the torque fluctuations of the form
\begin{equation}
	\langle \tau^2\rangle = \frac{1}{A}\langle F_i^{(L)}F_j^{(L)}\rangle I_{ij},
\end{equation}
where we have used the summation  convention and the moment of inertia tensor definition as
\begin{equation}
I_{ij} = \int_{S_1} d{\bf r}\ (\delta_{ij} r^2 - r_i r_j).
\end{equation}
The torque fluctuations are thus connected with the lateral force fluctuations through a geometric factor encoded by moment of inertia tensor. This result is completely general and valid for the assumed plane-parallel arrangement of the two disorder-carrying dielectric surfaces.

\subsection{General scaling of torque fluctuations}

We can now proceed to a general analysis of the torque fluctuations in terms of the area of the interacting surfaces $A$ as well as the normal separation between them $l$. As the lateral force fluctuations have the following scaling with respect to $A$ and $l$ \cite{dean2011}
\begin{equation}
\langle F_i^{(L)}F_j^{(L)}\rangle \sim \frac{A}{l^2},
\end{equation}
it follows that typical  torque fluctuations (or its room-mean-square $\tau_{\mathrm{rms}}$) scale as 
\begin{equation}
	\tau_{\mathrm{rms}} \sim  \frac{A}{l}, 
\end{equation}
which is thus {\em extensive} in terms of the area $A$. In other words, the magnitude of the variance of the random torque  exhibits a non-extensive scaling with area. This is  as one might expect,  because torque is
determined by the geometry of the area $A$ even in the limit of large area  (which is not the case for the random lateral force  \cite{dean2011}). However the geometry dependence of the torque fluctuations and the scaling with the area $A$ are obtained simply from the moment of inertia tensor.
For instance, if  instead of a disc-shaped area, one considers a square of area $A$ on the surface $S_1$
which can rotate about its center, then one finds that
$K_{jp} =\delta_{jp}A^2/12$
and consequently
\begin{equation}
	\langle \tau^2\rangle = \frac{A^2 g_{1s}g_{2s}}{24\pi}\int d k\,  k^3
	     [\tilde G(k;0,l)]^2 \tilde C_2( k)  \tilde C_1( k). 
\end{equation}
The torque fluctuations for  a square are thus slightly larger than that for a disc of the same area as one would intuitively expect.

\subsection{Torque fluctuations in a homogeneous system}

In the case where there are no dielectric jumps at the boundaries ($\varepsilon_1=\varepsilon_2=\varepsilon_m$) and the charge disorder 
is uncorrelated $C_\alpha({\bf r}-{\bf r}') = \delta({\bf r}-{\bf r}')$, 
the  expression (\ref{eq:pp_tau2}) reduces to
\begin{equation}
	\langle \tau^2\rangle = \frac{g_{1s}g_{2s} A^2}{128\pi^2 \varepsilon_0^2\varepsilon_m^2 l^2}. 
\end{equation}
This result can be obtained in a direct, more geometric manner, by starting from the variation in the bare Coulomb energy upon rotating the surface $S_1$ by a small angle around its
central axis, 
 \begin{equation}
 \delta E = \frac{1}{4\pi \varepsilon_0\varepsilon_m}\int d{\bf x}\,d{\bf y} \frac{\delta \rho_1({\bf x}) \rho_2({\bf y})}{|{\bf x} - {\bf y}|}, 
\end{equation}
and showing after some manipulations that
\begin{equation}
\label{eq:sxr_y}
  \langle \tau^2\rangle  \!= \! \frac{g_{1s}g_{2s} }{(4\pi \varepsilon_0\varepsilon_m)^2} \!\int\! d{\mathbf s} \, d{\mathbf r} \, \frac{(s_x r_y - s_y r_x)^2}{[({\mathbf r}-{\mathbf s})^2+l^2]^3}
    = \frac{g_{1s}g_{2s} A^2}{128\pi^2 \varepsilon_0^2\varepsilon_m^2 l^2}. 
\end{equation}

\begin{figure}[t]
\begin{center}
\vspace{-4cm}
\centerline{\includegraphics[angle=0,width=8cm]{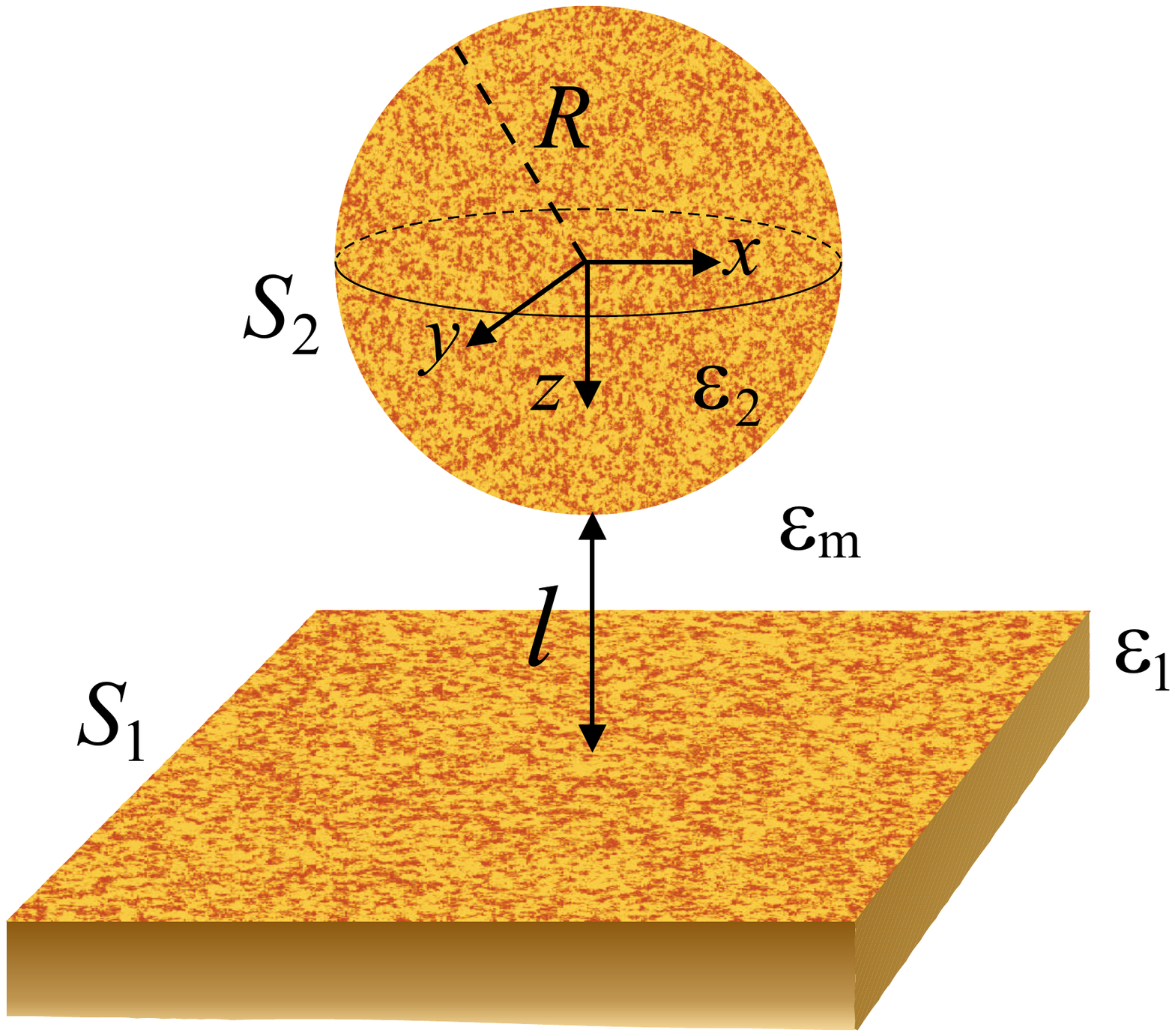}}
\vspace{-1cm}
\caption{(Color online) Schematic picture of a sphere ($S_2$) apposed to a planar substrate ($S_1$) at a minimum separation 
of $l$. Both objects carry random charges on their bounding surfaces  
(shown by small dark and light brown patches). The sphere is allowed to rotate around the central $z$ axis.}
\label{fig:fig_sp_schematic}
\vspace{-1cm}
\end{center}
\end{figure}

\section{Torque fluctuations in the sphere-plane geometry}

Let us now consider a sphere ($S_2$) of radius $R$ apposed to a planar substrate ($S_1$)
at a minimum separation of $l$ (see Fig. \ref{fig:fig_sp_schematic}), both carrying quenched random charges of density (variance) 
$g_{1s}$ and $g_{2s}$ over their surfaces only. 
The substrate  is assumed to be fixed and the sphere is assumed to be mounted on an axle that allows for rotation around the $z$ axis. 

We shall first focus on the case where there are no dielectric inhomogeneities at the boundaries ($\varepsilon_1=\varepsilon_2=\varepsilon_m$) and that  the charge disorder is 
 {\em uncorrelated}. The more general case of inhomogeneous dielectrics will be considered later (see below). 
In the absence of image charges, the sample-to-sample torque fluctuations in this system can be evaluated exactly as follows. 

The variation in the charge distribution of the sphere due a  small rotation $\delta \varphi$ around its central $z$ axis can be written as
  $\delta \rho_2({\mathbf x}) = - \delta \varphi \frac{\partial}{\partial \varphi} \rho_2({\mathbf x})$, 
  where ${\mathbf x}=(x, \theta, \varphi)$ and
\begin{equation}
  \rho_2({\mathbf x}) =  \sum_{n\in S_2} q_n \left(\frac{1}{x^2\sin \theta}\right) \delta(x-x_n) \, \delta(\theta-\theta_n)\, \delta(\varphi - \varphi_n)
\end{equation}
is the sphere charge distribution made up of point charges located at positions ${\mathbf x}_n$.
The variance of the energy of the system thus follows as
\begin{equation}
\langle (\delta E)^2\rangle = \int d{\bf x}\,d{\bf x}'\,d{\bf y}\,d{\bf y}'\frac{\langle \rho_1({\bf x})\rho_1({\bf x}')\rangle \langle \delta \rho_2({\bf y}) \delta \rho_2({\bf y}')\rangle}{|{\bf x} - {\bf y}||{\bf x}' - {\bf y}'|}, 
\end{equation}
where for uncorrelated charge disorder assumed here
\begin{eqnarray}
&&\!\!\!\!\langle \rho_1({\bf x})\rho_1({\bf x}')\rangle = g_{1s}\delta({\bf r}-{\bf r}')\delta(z-l-R)\delta(z'-l-R)\quad\,\,\,\,\,  \\
&&\!\!\!\!\langle  \delta\rho_2({\bf y}) \delta\rho_2({\bf y}')\rangle  = (\delta \varphi)^2\left(\frac{g_{2s}}{R^2\sin \theta}\right)\delta(y-R)\delta(y'-R)\nonumber\\
   &&\quad\quad\qquad\qquad\qquad\qquad\times\delta(\theta-\theta')\frac{\partial^2}{\partial\varphi\partial\varphi'} \delta(\varphi-\varphi'). 
\end{eqnarray}
Here we have introduced  ${\bf x}=({\bf r},z)\in S_1$ and ${\bf y}=(y, \theta, \varphi)\in S_2$. Then using $\langle (\delta E)^2\rangle = (\delta \varphi)^2\langle\tau^2\rangle$, we find
an expression resembling  Eq. (\ref{eq:sxr_y}), i.e.
\begin{eqnarray}
 && \langle \tau^2\rangle =  
			\frac{g_{1s}g_{2s} }{(4\pi \varepsilon_0\varepsilon_m)^2} \int d{\bf r} \, R^2\sin\theta\,  d\theta\, d\varphi\qquad\qquad\qquad\nonumber\\
		&&\qquad\qquad\quad\quad\times\frac{(s_x r_y - s_y r_x)^2}{\left[({\bf r} - {\bf s})^2
  				+(l+R(1-\cos\theta))^2\right]^3}, 
\end{eqnarray}
where ${\bf s} = (R\sin\theta\cos\varphi, R\sin\theta\sin\varphi)$, 
or explicitly, via a parameterization of ${\bf r}$ as ${\bf r} = (r\cos\alpha, r\sin\alpha)$, we find
\begin{figure*}[t!]
\begin{equation}
  \langle \tau^2\rangle
        = \frac{2\pi g_{1s}g_{2s} }{(4\pi \varepsilon_0\varepsilon_m)^2} \int_0^\infty \!dr \int_0^{\pi}\!d\theta \int_0^{2\pi}\!d\varphi \, 
       \frac{R^4 r^3\sin^3\theta\sin^2\varphi}{\left[R^2+r^2+(l+R)^2-2rR\sin\theta\cos\varphi-2R(l+R)\cos\theta\right]^3}
       \label{eq:tau2_sp}
\end{equation}
\hrule
\end{figure*}
\begin{center}
{\em see equation~(\ref{eq:tau2_sp})}
\end{center}
This equation can be written as  
\begin{equation}
\langle \tau^2\rangle = \frac{2\pi R^2 g_{1s}g_{2s} }{(4\pi \varepsilon_0\varepsilon_m)^2} f\left(\frac{l}{R}\right),
\end{equation} 
where the dimensionless function $f(v)$ reads
\begin{figure*}[t!]
\begin{equation}
  f(v)\equiv \int_0^\infty \!du \int_0^{\pi}\!d\theta \int_0^{2\pi}\!d\varphi \, 
       \frac{u^3\sin^3\theta\sin^2\varphi}{\left[1+u^2+(v+1)^2-2u\sin\theta\cos\varphi-2(v+1)\cos\theta\right]^3}.
       \label{eq:f_v}
\end{equation}
\hrule
\end{figure*}
\begin{center}
{\em see equation~(\ref{eq:f_v})}
\end{center}

Note that the same expression can be obtained if one assumes that the sphere $S_2$ is fixed and the substrate $S_1$ is allowed to freely rotate 
around the $z$ axis passing through the center of the sphere. 

\begin{figure}[t]
\begin{center}
\vspace{-5cm}
\centerline{\includegraphics[angle=0,width=8.5cm]{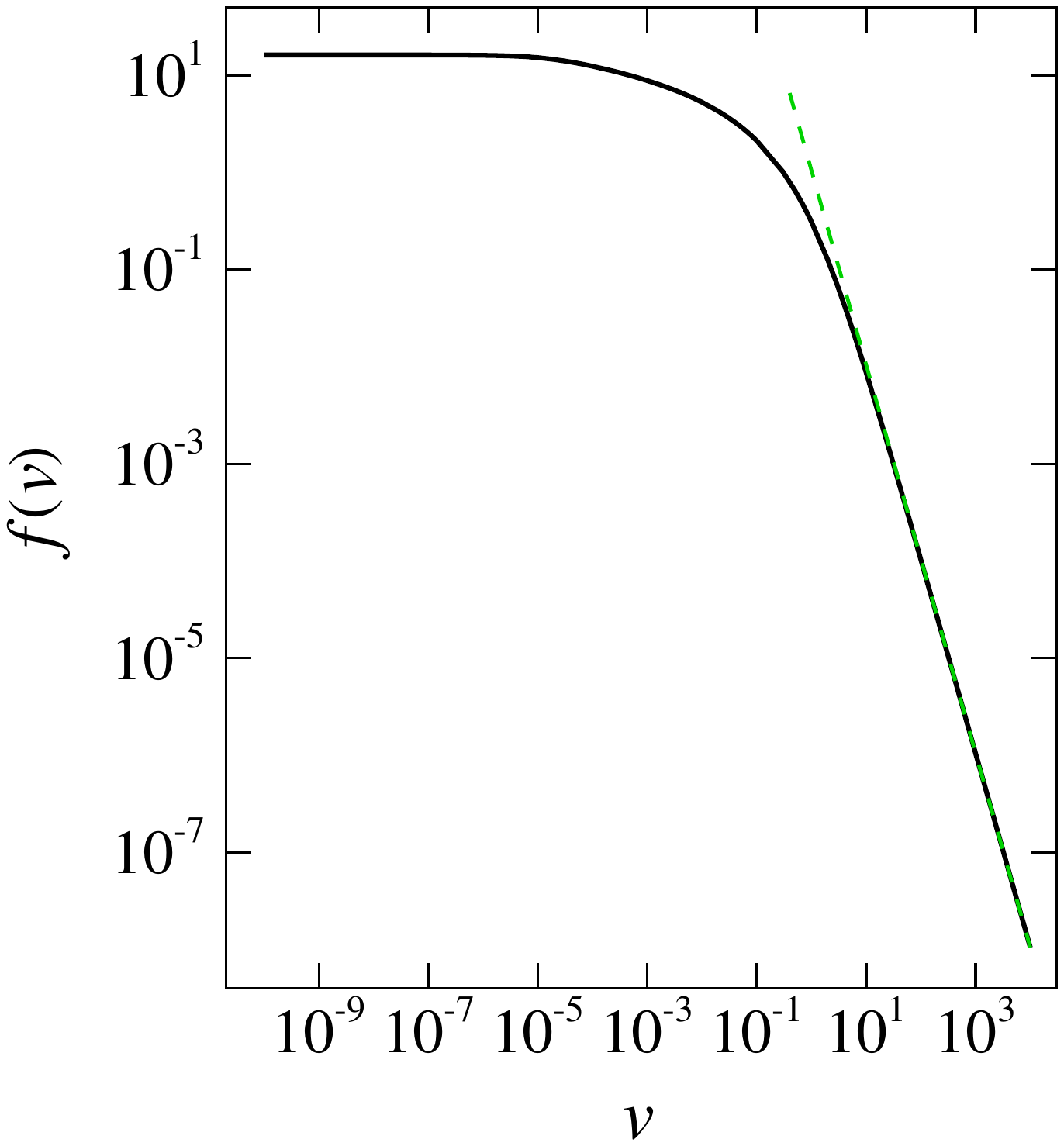}}
\vspace{-1cm}
\caption{(Color online) The log-log plot of the dimensionless function $f(v)$ as a function of the dimensionless sphere-substrate separation $v=l/R$ (solid line). The dashed (green) line shows the limiting behavior $f(v)\sim v^{-2}$ for large $v$.}
\label{fig:fig_f}
\end{center}
\end{figure}

In Fig. \ref{fig:fig_f}, we plot the dimensionless function $f(v)$ as a function of $v=l/R$. For large separation or small sphere radius $v=l/R\gg 1$,  the torque 
variance tends to zero and we find that $f(v)\simeq (1.0470\ldots)/v^2$; 
  hence
\begin{equation}
  \langle \tau^2\rangle \sim \frac{g_{1s}g_{2s} R^4  }{(4\pi \varepsilon_0\varepsilon_m)^2l^2}, 
\end{equation}
whereas for small separation or large sphere radius $v=l/R\ll 1$, we find  $f(v)\sim {\mathrm{const.}}\simeq 2\pi(16.1705\ldots)$;
hence
\begin{equation}
  \langle \tau^2\rangle \sim \frac{ g_{1s}g_{2s} R^2}{(4\pi \varepsilon_0\varepsilon_m)^2}.
\end{equation}

The dielectrically inhomogeneous case where the dielectric constants of the sphere and the substrate are in general different from $\varepsilon_m$ and
from each other is not tractable analytically. It is nevertheless possible to provide an estimate for the torque fluctuations variance in analogy with the
results presented in Ref. \cite{dean2011} for the lateral force fluctuations in the dielectric sphere-substrate system. There it was shown, using the proximity force
arguments, that the dielectric effects lead only to a correction of the prefactor of the result obtained for a homogeneous sphere-substrate system, where
the dielectric-dependent prefactor is exactly the same as obtained for two semi-infinite planar slabs \cite{dean2011}. 

This result then suggests the following heuristic estimate for the torque variance in the dielectrically inhomogeneous sphere-substrate system 
\begin{equation}
  \langle \tau^2\rangle = \frac{2 R^2 g_{1s}g_{2s} }{\pi \varepsilon_0^2} \left[\frac{\varepsilon_m^2|\ln(1-\Delta_1\Delta_2)|}{(\varepsilon_m+\varepsilon_1)^2(\varepsilon_m+\varepsilon_2)^2\Delta_1\Delta_2}\right]
         f\left(\frac{l}{R}\right), 
\end{equation}
 where
   $ \Delta_\alpha = \frac{\varepsilon_\alpha - \varepsilon_m}{\varepsilon_\alpha+\varepsilon_m}$ for $\alpha=1,2$.
The above result obviously reduces to the one in Eq. (\ref{eq:tau2_sp}) when  $\varepsilon_1=\varepsilon_2=\varepsilon_m$.

\section{Discussion and Conclusions}

Generalizing our previous results on the force fluctuations between charge-disordered interfaces, we have now
calculated fluctuations in the torque between two coaxial randomly charged surfaces. Surprisingly, the disorder-induced torque fluctuations scale differently with the surface area of the interfaces bearing charge disorder, with the root-mean-square torque being {\em extensive}
in the surface area $A$. In our opinion this opens up a feasible way to measure charge disorder-induced interactions between randomly charged media in a way which is independent of normal force 
measurements and with a higher signal than lateral force measurements.

The measurements we have in mind would of course have to be sample-to-sample torque fluctuations. While they decay with the separation between the bounding surfaces carrying the disordered charges in a similar way as the forces, the magnitude of torque fluctuations is $\sqrt{A}$ times larger than the corresponding lateral or normal force fluctuations which are always comparable in magnitude. Results obtained for the sphere-plane geometry in fact indicate that the disorder effect on the torque fluctuations should be highest in this case, asymptotically leveling off at a finite value for small separations. 

That vdW interactions between anisotropic media lead in general to torques induced by electromagnetic field fluctuations was first realized by Weiss and Parsegian \cite{Weiss,measuretorque}. The effect persists not only in the non-retarded but also in the retarded limit \cite{Barash} and should be experimentally detectable with modern instrumentation \cite{Chen} such as  the torsion-balance-based setups \cite{torsion}. It is worth stressing that in the case of vdW torques the fluctuations are  in the field mediating the interactions, while the boundaries do not fluctuate and are not disordered. In the case analyzed here however, the effect is very different: the field does not fluctuate but its sources indeed are quenched in a statistically disordered state. The other important difference is that  vdW torques  arise due to the non-isotropic dielectric response of the interacting bodies, while the disorder-generated torque fluctuations are present even between bodies with a completely isotropic--but disordered--charge distribution, being in this sense obviously more universal.

In order to detect sample-to-sample variation in the torque or the force itself, one would have to perform many experiments  measuring the value of the  torque as one object is gradually rotated above the other. If the measurement of the torque at any given angle can be done to an accuracy greater than
the value of the torque fluctuations estimated here, then valuable information about the effect of charge disorder induced interactions could be extracted.

\begin{acknowledgement}
A.N. acknowledges support from the Royal Society, the Royal Academy of Engineering, and the British Academy and useful discussions with M.M. Sheikh-Jabbari. D.S.D. acknowledges support from the Institut Universitaire de France. R.P. acknowledges support from ARRS through research program P1-0055 and research project J1-0908 as well as from the University of Toulouse for a one month position of {\em Professeur invit\' e} at the Department of Physics. 
\end{acknowledgement}


\end{document}